\documentclass[twocolumn]{aastex63}
\usepackage{lineno}
\usepackage{epstopdf}
\usepackage{comment}
\usepackage{amsmath}
\shorttitle{Chromospheric Evaporation of M-dwarf}
\shortauthors{Wang et al.}

\begin{document}

\title{Potential Chromospheric Evaporation in A M-dwarf's Flare Triggered by Einstein Probe Mission}

\correspondingauthor{J. Wang, H.Y. Liu}
\email{wj@nao.cas.cn, liuheyang@nao.cas.cn}

\author{J. Wang}
\affiliation{Key Laboratory of Space Astronomy and Technology, National Astronomical Observatories, Chinese Academy of Sciences, Beijing 100101,
People's Republic of China}
\affiliation{Guangxi Key Laboratory for Relativistic Astrophysics, School of Physical Science and Technology, Guangxi University, Nanning 530004,
People's Republic of China}
\affiliation{GXU-NAOC Center for Astrophysics and Space Sciences, Nanning, 530004, People's Republic of China}

\author{X. Mao}
\affiliation{Key Laboratory of Space Astronomy and Technology, National Astronomical Observatories, Chinese Academy of Sciences, Beijing 100101,
People's Republic of China}
\affiliation{School of Astronomy and Space Science, University of Chinese Academy of Sciences, Beijing, People's Republic of China}

\author{C. Gao}
\affiliation{Guangxi Key Laboratory for Relativistic Astrophysics, School of Physical Science and Technology, Guangxi University, Nanning 530004,
People's Republic of China}
\affiliation{Key Laboratory of Space Astronomy and Technology, National Astronomical Observatories, Chinese Academy of Sciences, Beijing 100101,
People's Republic of China}
\affiliation{School of Astronomy and Space Science, University of Chinese Academy of Sciences, Beijing, People's Republic of China}

\author{H. Y. Liu}
\affiliation{Key Laboratory of Space Astronomy and Technology, National Astronomical Observatories, Chinese Academy of Sciences, Beijing 100101,
People's Republic of China}

\author{H. L. Li}
\affiliation{Key Laboratory of Space Astronomy and Technology, National Astronomical Observatories, Chinese Academy of Sciences, Beijing 100101,
People's Republic of China}

\author{H. W. Pan}
\affiliation{Key Laboratory of Space Astronomy and Technology, National Astronomical Observatories, Chinese Academy of Sciences, Beijing 100101,
People's Republic of China}

\author{C. Wu}
\affiliation{Key Laboratory of Space Astronomy and Technology, National Astronomical Observatories, Chinese Academy of Sciences, Beijing 100101,
People's Republic of China}
\affiliation{School of Astronomy and Space Science, University of Chinese Academy of Sciences, Beijing, People's Republic of China}

\author{Y. Liu}
\affiliation{Key Laboratory of Space Astronomy and Technology, National Astronomical Observatories, Chinese Academy of Sciences, Beijing 100101,
People's Republic of China}

\author{G. W. Li}
\affiliation{Key Laboratory of Space Astronomy and Technology, National Astronomical Observatories, Chinese Academy of Sciences, Beijing 100101,
People's Republic of China}

\author{L. P. Xin}
\affiliation{Key Laboratory of Space Astronomy and Technology, National Astronomical Observatories, Chinese Academy of Sciences, Beijing 100101,
People's Republic of China}

\author{S. Jin}
\affiliation{Guangxi Key Laboratory for Relativistic Astrophysics, School of Physical Science and Technology, Guangxi University, Nanning 530004,
People's Republic of China}
\affiliation{Key Laboratory of Space Astronomy and Technology, National Astronomical Observatories, Chinese Academy of Sciences, Beijing 100101,
People's Republic of China}
\affiliation{School of Astronomy and Space Science, University of Chinese Academy of Sciences, Beijing, People's Republic of China}

\author{D. W. Xu}
\affiliation{Key Laboratory of Space Astronomy and Technology, National Astronomical Observatories, Chinese Academy of Sciences, Beijing 100101,
People's Republic of China}
\affiliation{School of Astronomy and Space Science, University of Chinese Academy of Sciences, Beijing, People's Republic of China}

\author{E. W. Liang}
\affiliation{Guangxi Key Laboratory for Relativistic Astrophysics, School of Physical Science and Technology, Guangxi University, Nanning 530004,
People's Republic of China}
\affiliation{GXU-NAOC Center for Astrophysics and Space Sciences, Nanning, 530004, People's Republic of China}

\author{W. M. Yuan}
\affiliation{Key Laboratory of Space Astronomy and Technology, National Astronomical Observatories, Chinese Academy of Sciences, Beijing 100101,
People's Republic of China}

\author{J. Y. Wei}
\affiliation{Key Laboratory of Space Astronomy and Technology, National Astronomical Observatories, Chinese Academy of Sciences, Beijing 100101,
People's Republic of China}
\affiliation{School of Astronomy and Space Science, University of Chinese Academy of Sciences, Beijing, People's Republic of China}






\begin{abstract}
Although flares from late-type main-sequence stars have been frequently detected in 
multi-wavelength, the associated dynamical process has been rarely reported so far.
Here, we report follow-up observations of 
an X-ray transient triggered by WXT onboard the Einstein Probe at 
UT08:45:08 in 2024, May 7. The photometry in multi-bands and time-resolved spectroscopy
started at 3 and 7.5 hours after the trigger, respectively, which enables \rm us to identify the transient
as a flare of the M-dwarf 2MASS~J12184187-0609123. The bolometric energy released in the flare is estimated to be $\sim10^{36}\ \mathrm{erg}$ from its X-ray light curve. The H$\alpha$ emission-line
profile obtained at about 7 hours after the trigger shows an evident 
blue asymmetry with a maximum velocity of $200-250\ \mathrm{km\ s^{-1}}$.
The blue wing can be likely explained by the chromospheric temperature (cool) upflow associated
with chromospheric evaporation, in which 
the mass of the evaporating plasma is estimated to be $1.2\times10^{18}$g. 
In addition, a prominence eruption with an estimated mass of 
$7\times10^{15}\mathrm{g}<M_{\mathrm{p}}<7\times10^{18}\mathrm{g}$ can not be 
entirely excluded.
\end{abstract}



\section{Introduction} \label{sec:intro}

The habitability of an exoplanet is predicted to depend on the activity of the host 
star (e.g., Cherenkov et al. 2017; Garcia-Sage et al. 2017; Airapetian et al. 2016, 2017;
Tian et al. 2011; Barnes et al. 2016; Chen et al. 2021). 
By analogizing with the Sun, the 
activity is believed to be resulted from stellar magnetic reconnection (e.g., Noyes et al. 1984;
Wright et al. 2011; Shulyak et al. 2017), although the process that creates and maintains 
these magnetic fields is still an open issue in late-type stars because of a lack of the  
boundary between the radiative and convective zones (e.g., Hotta et al. 2022; Bhatia et al. 2023).

The magnetic reconnection not only triggers highly energetic flares that have  
been frequently observed in a fraction of late type stars in multi-wavelength from X-ray 
to radio (e.g., Pettersen 1989; Schmitt 1994; Osten et al. 2004, 2005; Huenemoerder et al.
2010; Maehara et al. 2012; Kowalski et al. 2013; Balona 2015;
Davenport et al. 2016; Notsu et al. 2016; Van Doorsselaere
et al. 2017; Chang et al. 2018; Paudel et al. 2018; Schmidt et al.
2019; Xin et al. 2021, 2024; Li et al. 2023b, Li et al. 2023a, 2024; Bai et al. 2023), 
but also produces complicate dynamics, including a coronal mass ejection (CME),
i.e., a large scale expulsion of the confined plasma into interplanetary space along an open magnetic 
field, and a
upward (evaporation) or downward (condensation) expansion with a velocity of 
$10^{1-2}\ \mathrm{km\ s^{-1}}$ when 
the chromospheric plasma is overpressured.

In contrast to the flares, the detection of the aforementioned dynamics 
is still a hard task for distant stars because of the insufficient spatial resolution of
contemporary instruments. We refer the readers to Leitzinger \& Odert (2022) for a 
recent review on this issue. Briefly speaking, there are, so far, 
only $\sim50$ CME candidates detected 
through different methods (e.g., Wang et al. 2021, 2022; Wang 2023; Namekata et al. 2021; Argiroffi et al. 2019;
Veronig et al. 2021; Lu et al. 2022; Maehara et al. 2021; Chen et al. 2022;
Inoue et al. 2023, 2024a, 2024b; Notsu et al. 2024; Namekata et al. 2024). 
Cao \& Gu (2024) recently reported a possible CME based on a detection of 
significant red  asymmetry of the H$\alpha$ emission-line in active RS CVn star Pegasi. 
In addition, the possible chromospheric evaporation or condensation has been reported in 
a few stars according to their observed emission-line asymmetry or bulk velocity shift (e.g., 
Wu et al. 2022; Wang et al. 2022; Cao \& Gu 2024).

In this paper, we report an identification of a X-ray transient triggered by the Einstein Probe mission by
photometric and time-resolved spectroscopic follow-up observations in optical bands. 
Our follow-ups enable us to not only identify the 
event as a flare of a M-dwarf, but also detect a blue asymmetry of the H$\alpha$ 
emission-line profile, which is possibly caused by chromospheric evaporation.

The paper is organized as follows. Sections 2 presents the trigger by the Einstein Probe mission.
Our follow-up observations in optical bands, along with data reduction, are given in Section 3. 
Section 4 shows that results and analysis. The implications are presented in Section 5.

\section{X-ray Transients Detected by EP Mission} \label{sec:style}

An X-ray transient designated EP240507a was triggered by the Wide-field X-ray Telescope (WXT) on board Einstein Probe (EP) at 2024-05-07UT08:45:08 (MJD~$=$~60437.36468).  
The observations of this transient started at 2024-05-07UT08:14:10, and the light curve shows no significant variations within the first 3000~s, and decays gradually in the subsequent 5 hours. 
The WXT position is determined to be R.A.=184.\degr670 and DEC=$-6.\degr165$ with an uncertainty of 3\arcmin\ (radius, 90\%\ C.L. statistical and systematic). 
It is worth noting that another flare showing similar light curve occurring at 2024-05-04UT22:27:45 was also detected by EP-WXT, which is spatially consistent with EP240507a at a separation of 43\arcsec. 
This event was not triggered by on-board processing unit since its peak flux is lower than the threshold.
Within the localization error circle of both transients, there exists an eROSITA source, 1eRASS J121842.0-060913, with a flux of $6.17\times10^{-13}\ \mathrm{erg\ s^{-1}\ cm^{-2}}$ in 0.2--2.3 keV (Predehl et al. 2021; Merloni et al. 2024), which is lower than the peak flux detected by EP-WXT by two orders of magnitudes. 
This eROSITA source is separated by only 2.4\arcsec\ from a bright M-dwarf 2MASS~J12184187-0609123 (see Table~\ref{tab:properties} for its basic properties), indicating these two X-ray transients are likely stellar flares (named as Flare~1 and Flare~2) associated with this star.


\begin{deluxetable*}{ccccccc}
\centering
\tablecaption{Observation log and the best-fit spectral parameters for each EP/WXT observation\label{tab:ep_spec}}
\tablehead{
\colhead{Flare No.} & \colhead{Obs. Start time} & \colhead{Exposure time} & \colhead{$kT_{\mathrm{e}}$\tablenotemark{a}} & \colhead{$EM$\tablenotemark{b}} & \colhead{$L_\mathrm{X}$\tablenotemark{c}} & \colhead{$\chi^2_\nu$ (d.o.f.)\tablenotemark{d}}\\
\colhead{} & \colhead{(UTC)} & \colhead{(s)} & \colhead{keV} & \colhead{$10^{52}$ cm$^{-3}$}& \colhead{$10^{30}$ erg s$^{-1}$} & \colhead{}\\
\colhead{(1)} & \colhead{(2)} & \colhead{(3)} & \colhead{(4)} & \colhead{(5)} & \colhead{(6)} & \colhead{(7)}
} 
\startdata
1  & 2024-05-04 22:27:45 & 2012  & $>$4.7 & 37.9$_{-5.9}^{+5.2}$       & 3.7$_{-0.6}^{+0.9}$\tablenotemark{\textnormal{*}}    & 0.96 (52) \\
1  & 2024-05-05 00:04:04 & 2284  & $>$4.5 & 16.3$_{-4.1}^{+3.1}$       & 1.5$_{-0.4}^{+0.5}$     & 0.94 (29) \\
1  & 2024-05-05 01:40:24 & 3084  & 1.2$_{-0.2}^{+0.2}$ & 2.3$_{-0.8}^{+1.4}$       & 0.5$_{-0.1}^{+0.2}$     & 0.92 (34) \\
2  & 2024-05-07 08:14:10 & 3090  & $>$3.9 & 41.1$_{-5.6}^{+8.4}$       & 5.7$_{-0.7}^{+0.5}$\tablenotemark{\textnormal{*}}     & 0.88 (84) \\
2  & 2024-05-07 09:50:29 & 3090  & $>$3.8 & 21.8$_{-5.5}^{+3.0}$        & 2.2$_{-1.1}^{+0.8}$     & 1.12 (51) \\
2  & 2024-05-07 11:26:49 & 3090  & $>$5.2 & 17.1$_{-3.4}^{+2.6}$       & 1.9$\pm0.5$     & 1.13 (41) \\
2  & 2024-05-07 13:03:08 & 3091  & 1.6$_{-0.4}^{+4.3}$ & 3.3$_{-1.8}^{+3.6}$       & 0.5$\pm0.2$     & 1.36 (26)\\
\enddata
\tablecomments{All the errors are at the 1$\sigma$ significance level and are determined by the task of \tt error \rm in the Xspec package.\rm}
\tablenotetext{a}{Plasma temperature.}
\tablenotetext{b}{The EM of Plasma, calculated as $EM=4\pi d^2\times 10^{14}\times norm$, where $norm$ is one of the parameters in the {\tt\string apec} model.}
\tablenotetext{c}{The flare luminosity in 0.5--4.0 keV, calculated by $L_\mathrm{X}=4\pi d^2F_\mathrm{X}$, where $d$ is the distance of 2MASS J12184187-0609123 and $F_\mathrm{X}$ is the flare flux, assumed to be isotropic.}
\tablenotetext{d}{Reduced $\chi^2$ along with the degrees of freedom.}
\tablenotetext{*}{The detected peak X-ray luminosity of Flare 1 and 2.}
\end{deluxetable*}

\begin{figure*}
    \centering
    \includegraphics[width=0.8\linewidth]{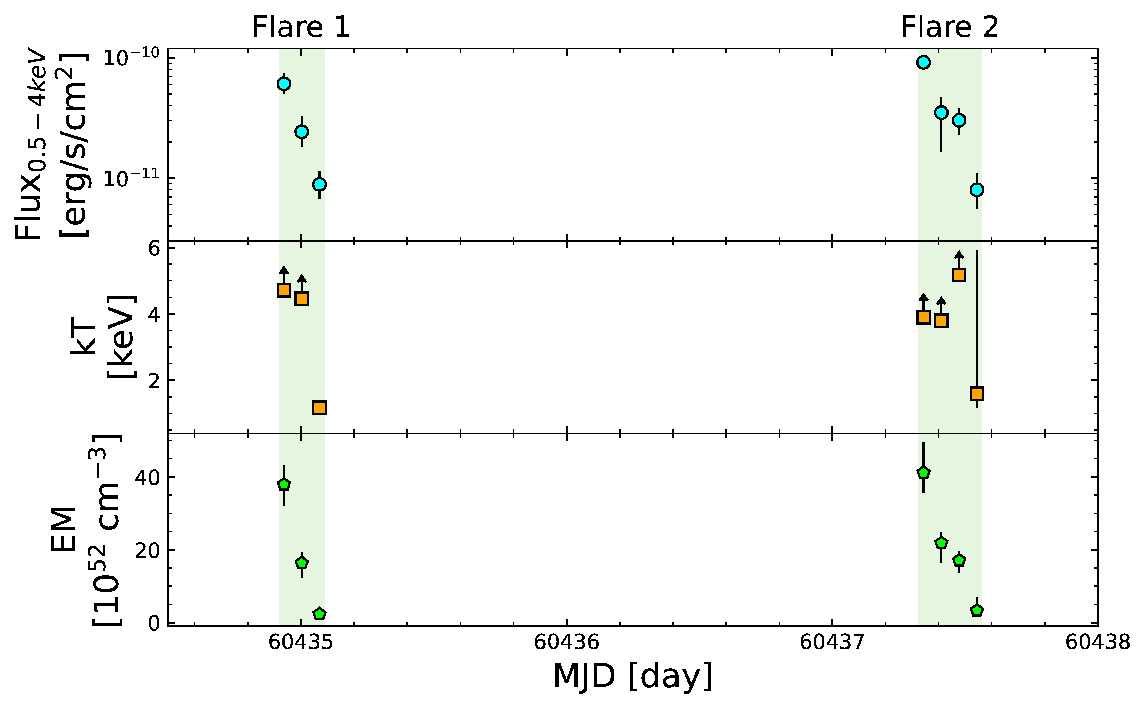}
    \caption{The X-ray light curves and temporal evolution of the spectral parameters measured from the X-ray spectra of the two transients detected by EP/WXT. The top panel shows the light curves in 0.5--4.0 KeV. The middle and bottom panels present the temporal variations of temperature and emission measure of the plasma component, respectively. All the errors plotted in the figure are at a 
    significance level of 1$\sigma$.  \label{fig:xlcurve_para}}
\end{figure*}

\begin{table}
        \centering
        \caption{Properties of 2MASS~J12184187-0609123}
        \label{tab:example_table}
        \begin{tabular}{cc} 
        \hline
        \hline
        Property & Value \\
          (1) & (2)  \\     
        \hline     
       Gaia DR2 ID & 3584656640992625280 \\
       G-band (mag) & $13.205\pm0.003$\\
       GBp-GRp (mag) & $3.026\pm0.007$  \\
       B-band (mag) & 16.34\\
       R-band (mag) & 13.43\\
       Distance (pc) & $22.64\pm0.05$\\
       $M_G$ (mag) & 11.43\\
       S.p. Type  & M4\\
       $T_{\mathrm{eff}}$ (K) & $3168.0\pm157.0$\\
       $R_\star$ ($R_\odot$) & $0.256\pm0.008$\\
       $M_\star$ ($M_\odot$) & $0.227\pm0.020$\\ 
       $\log(g/\mathrm{cm\ s^{-2}})$ & $4.98\pm0.01$\\
        \hline
        \end{tabular}
        \tablecomments{References: Gaia Collaboration et al. (2022); Monet et al. (2003); Stassun et al. (2019)}
        \label{tab:properties}
\end{table}

\section{Optical Follow-up Observations and Data Reductions}

After the trigger, optical follow-up observations were carried out by multiple ground-based
telescopes to identify the optical counterpart of the X-ray
transients and to study its properties in both photometry and spectroscopy.  

\subsection{Follow-ups in Photometry}

The field of the X-ray transient was monitored by the Chinese Ground Follow-up Telescope (C-GFT)\footnote{C-GFT is deployed at 
Changchun observatory, National Astronomical Observatories, Chinese Academy of Sciences (NAOC), and has a diameter of 1.2m. The telescope works in the 
three SDSS $g$, $r$ and $i-$bands simultaneously.
By equipped with a $\mathrm{2k\times2k}$ CMOS-CCD in each channel, the $f$-ratio of 8 leads to a field-of-view of  
$21\times21\mathrm{arcmin^2}$. 
The sensitivity of C-GFT is typical of $r=19$mag (AB) at $5\sigma$ significance level for an exposure of 100s in dark night.} 
of the SVOM mission\footnote{SVOM, launched in 2024, June 22, is a Chinese-French space mission dedicated to the detection and
study of gamma-ray bursts. Please see Atteia et al. (2022) and the white paper
given by Wei et al. (2016) for details} in SDSS $g$, $r$ and $i$-bands simultaneously. The monitor 
started at MJD=60437.9866627 day, i.e., about 3 hours after the trigger. The exposure time
is 100 seconds for each frame.




Follow-ups in photometry about 7.5 hours after the trigger were additionally carried out by
the GWAC-F60A telescope (e.g., Han et al. 2021; Xu et al. 2020)
in the standard Johnson–Cousins $B$ and $R$-bands. The exposure 
time is 60 seconds for each frame. 

All the raw images were reduced by the IRAF package\footnote{IRAF is distributed by the National Optical Astronomical Observatories,
which are operated by the Association of Universities for Research in
Astronomy, Inc., under cooperative agreement with the National Science
Foundation.} 
through the standard routines, including bias and flat-field corrections.
The standard aperture photometry were applied to the location of 2MASS~J12184187-0609123.

The effect of reddening caused by the Galaxy is ignored in our photometry because of
the extremely small color excess $E(B-V)$. Given the determined distance of $d=22.64$pc,
a rough estimation of $E(B-V)=0.01$mag can be
inferred from the relationship $E(B-V)\approx0.53\times(d/\mathrm{kpc})$, where the hydrogen density around the Sun of $n_{\mathrm{H}} = 10^6\ \mathrm{cm^{-3}}$ and the constant dust-to-gas ratio are assumed (Bohlin et al. 1978). In addition, based on
the updated dust reddening map provided by Schlafly \& Finkbeiner (2011). 
The extinction in the Galactic plane along the line of
sight (LOS) of the object is determined to be as low as $E(B-V)=0.03$mag.


\subsection{Spectroscopy by NAOC 2.16m telescope}

We started a time-resolved long-slit spectroscopic monitor on the object 2MASS~J12184187-0609123  by the NAOC 2.16 m telescope 
(Fan et al. 2016) at about 7 hours after the trigger.
Eight spectra in total were obtained in the night of 2024 May 07, and one quiescent spectrum in 2024 June 01. 

All the spectra were taken
by the Beijing Faint Object Spectrograph and Camera
that is equipped with a back-illuminated E2V55-30 AIMO CCD. The exposure time of each frame 
is 220 seconds for the observation run in 2024 May 07, and 480 seconds for the 
quiescent spectrum.
The G8 grism with a wavelength coverage of 5800 to 8200\AA\ was used in the
observations, which allows us to study the emission-line profile of H$\alpha$ with 
adequate spectral resolution. With a slit width of 1.8\arcsec oriented in the 
south–north direction, the resolution is measured to be 3.5\AA\ 
according to the sky emission lines, 
which corresponds to $R=\lambda/\Delta\lambda=1880$ and a velocity of $160\ \mathrm{km\ s^{-1}}$ at the H$\alpha$ emission line. The wavelength calibration was carried out with iron–argon comparison
lamps. Flux calibration of all spectra was carried out with
observations of the standard stars from the Kitt Peak National Observatory (Massey et al. 1988).

After discarding the 7th spectrum obtained in 2024 May 07 due to its low signal-to-noise ratio, 
one-dimensional (1D) spectra were
extracted from the raw images by using the IRAF package
and standard procedures, including bias subtraction and flat-field correction.
The apertures of both the object and sky
emission were fixed in the spectral extraction of both the object and corresponding standard.
The extracted 1D spectra were
then calibrated in wavelength and in flux by the corresponding
comparison lamp and standard stars. The zero-point of the
wavelength calibration was corrected for each spectrum by using 
the sky [\ion{O}{1}]$\lambda$6300 emission line as a reference. 
The accuracy of wavelength calibration is therefore resulted to be $\sim0.1$\AA, 
which corresponds to a velocity of $\sim5\ \mathrm{km\ s^{-1}}$ for the H$\alpha$ line.

\section{Results and Analysis}

\subsection{X-ray Spectra and Light Curves}


The X-ray spectra is fitted using the Xspec software package and the {\tt\string apec}  model (e.g. Arnaud et al. 2010, Smith et al. 2001). The absorption and the metal abundance cannot be well constrained due to the limited counts. Hence these two parameters are fixed to 0 and 1.0$Z_{\odot}$ by default, respectively. The best-fit temperature $kT_{\mathrm{e}}$ and emission measurement $EM$ are presented in Table \ref{tab:ep_spec},  and their temporal evolution is shown in the middle and bottom panels in Figure \ref{fig:xlcurve_para}, respectively.
All the uncertainties 
are determined by the \tt error \rm task in the Xspec package.\rm  

Since only the decay phase was detected in both flares, the X-ray light curve of each flare is fitted by a simple exponential decay model:
\begin{equation}
c(t)=(c_\mathrm{p}-c_\mathrm{q})\times\exp\bigg(-\frac{t-t_\mathrm{p}}{\tau}\bigg)+c_\mathrm{q}, (t_\mathrm{p} \leq t),
\label{equ:counts_rate}
\end{equation}
where $t_\mathrm{p}$ and $\tau$ are the peak time and e-folding decay time, respectively. 
$c_\mathrm{p}$ and $c_\mathrm{q}$ are the measured flaring and quiescent photon fluxes, respectively. In each flare, 
both $c_\mathrm{p}$ and $t_\mathrm{p}$ are determined from the 
first detection.
$c_\mathrm{q}$ is estimated to be $3.3\times10^{-4}\ \mathrm{photons\ s^{-1}\ cm^{-2}}$, by transforming the 
eROSITA flux to the EP-WXT photon flux in 0.5-4 keV.

\begin{figure*}
    \centering
    \includegraphics[width=0.8\linewidth]{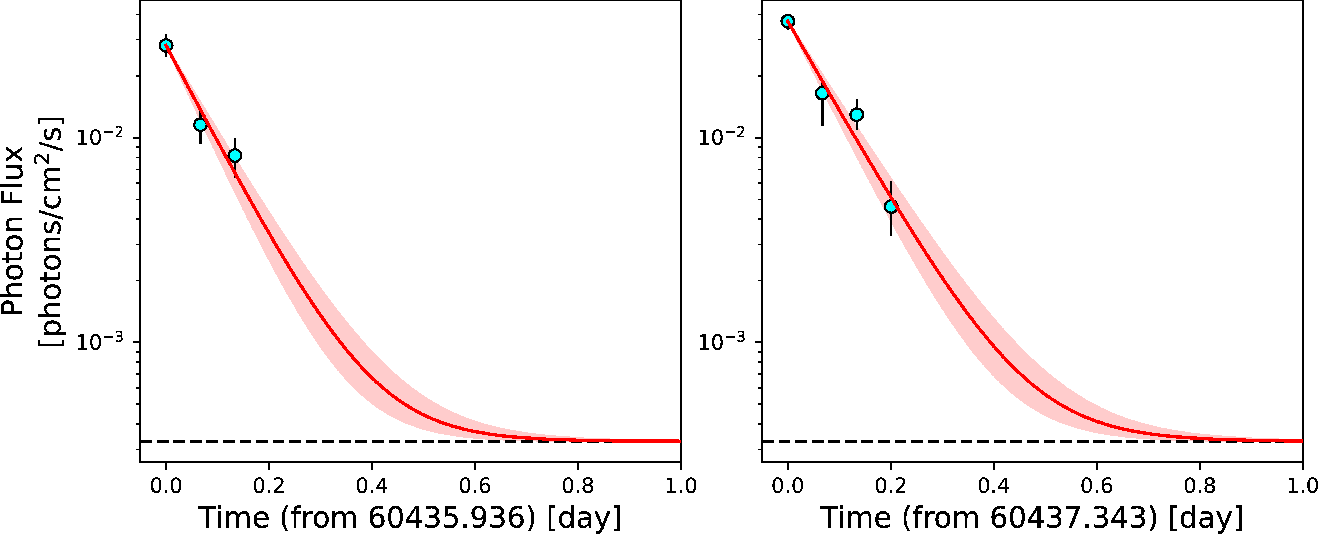}
    \caption{An illustration of the modeling for the X-ray light curves of the Flare 1 (left panel) and Flare 2 (right panel). The best-fit exponential decay models are over plotted by the red solid lines in both panels. The black dashed lines mark the corresponding
quiescent flux determined from the eROSITA survey.
    \label{fig:xlcurve_fit2}}
\end{figure*}

Figure \ref{fig:xlcurve_fit2} shows that both light curves can be well fitted by the model. The best-fit e-folding decay time is $\tau=7.86\pm1.01$ ks for the Flare 1 and $\tau=8.50\pm0.83$ ks for the Flare 2. 
The flaring energy released in the EP/WXT band 
is estimated to be $E_\mathrm{X}=\tau L_\mathrm{X,p}=2.9_{-0.6}^{+0.8}\times 10^{34}$ erg and 4.8$_{-0.8}^{+0.6} \times 10^{34}$ erg for the 
Flare 1 and 2, respectively, indicating these two events are both superflares.


\begin{figure*}
    \centering
    \includegraphics[width=0.5\linewidth]{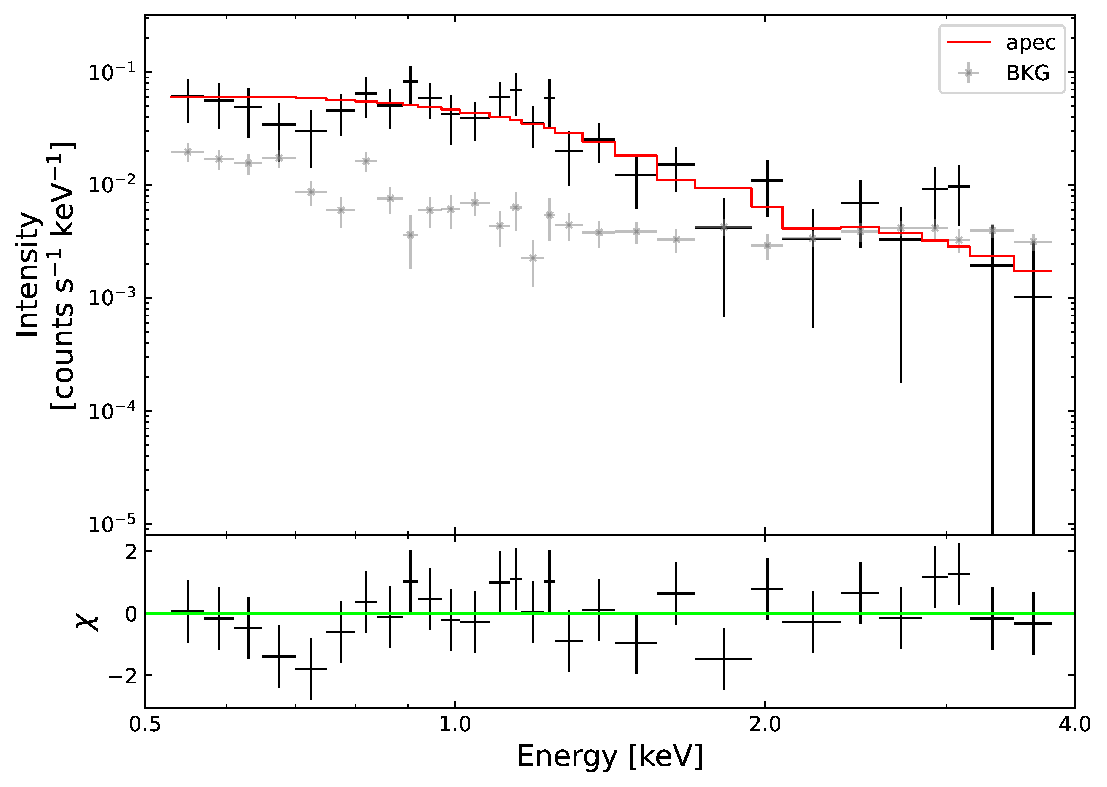}
    \caption{Illustration for the fitting of the X-ray spectrum of Flare 2 observed at  2024-05-07UT08:14:10 with the {\tt\string apec} model. The bottom panel underneath the spectrum shows the deviations.
    \label{fig:xspec_fit}}
\end{figure*}

\subsection{Multi-bands Light Curves}

The multi-bands light curves of the object 2MASS~J12184187-0609123 obtained in 2024, 
May 7th are
presented in Figure 4, in which we transfer the Johnson $B-$ and $R-$band magnitudes to 
the SDSS $g-$ and $r-$band magnitudes, respectively, according to the photometry of the 
field stars. The 
accuracy of this transformation is determined to be 0.06 and 0.01 mag in the 
$B-$ and $R-$bands, respectively.

As revealed by the C-GFT, one can see from the figure that the object shows
a gradually decreased brightness in both $g$ and $r$-bands. Combining the light curves
with the excess of H$\alpha$ line emission (see below) strongly indicates that both X-ray transients result from the stellar flare activity of a M4 red dwarf. Compared with
the quiescent brightness of object\footnote{The quiescent brightness in the SDSS photometry system is 
determined to be $g_{\mathrm{SDSS}}=15.6097\pm0.0833$, $r_{\mathrm{SDSS}}=14.1538\pm0.0110$ and $i_{\mathrm{SDSS}}=12.5389\pm0.02736$ mag from  both the Pan-STARRS 
Survey Data Release 1 (Chambers et al. 2016) and the transformation given in Tonry et al. (2012).}, the continuum emission of the flare almost fades out at $\sim7.5$ hr after the trigger, as revealed by our late monitor carried out by the F60A
telescope, although this is not true for the H$\alpha$ line emission as shown below.

We then model the $r-$band light curve by a simple exponential decay $F=F_{\mathrm{p}}e^{-(t-t_0)/\tau}$ by following
previous studies,  
where $F$ and $F_{\mathrm{p}}$ are relative flux 
normalized to the quiescent level for the decay and peak, 
$\tau$ the e-folding decay time, $t_0$ the peak time that is fixed to be the 
EP/WXT trigger time. The modeling is shown
in the inset panel in Figure 4. Based on the fitting,
the equivalent duration time ($ED$), i.e., the time needed
to emit the flare energy at the quiescent level, is estimated to be 
$\approx1.7\times10^4$s or $\sim4.8$hr, which leads to an energy 
release in the $g-$band of $E_g=4\pi d^{2}\times ED\times F_0 = 1.1\times10^{31}$erg. 
One should bear in mind that this value is a lower limit because multiple components are 
needed to reproduce the observed optical light curve of a flare at early epoch
(e.g., Davenport et al. 2014; Xin et al. 2021, 2024; 
Wang et al. 2021, 2022; Li et al. 2023), which is 
however not covered by our follow-ups in optical bands.

\begin{figure*}[ht!]
\plotone{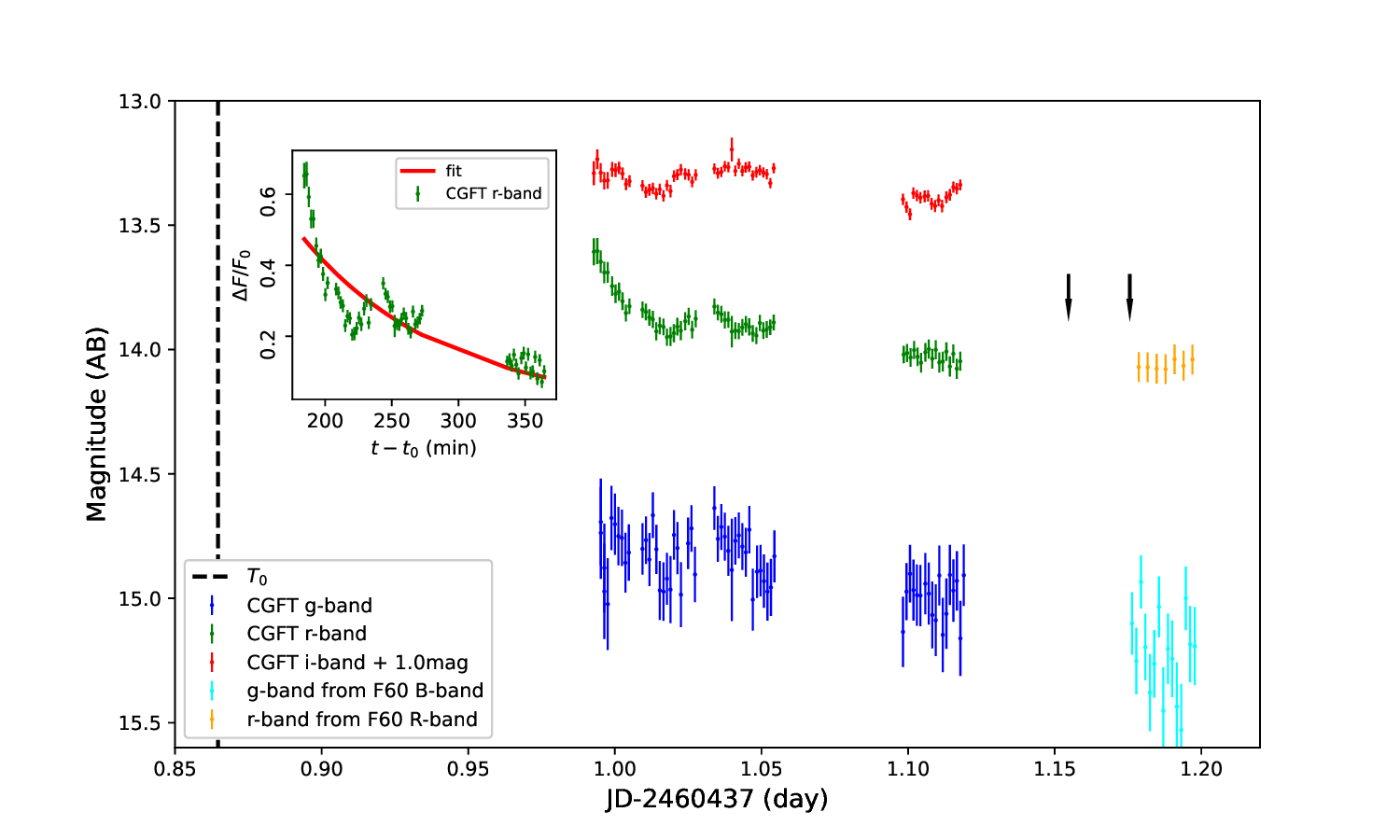}
\caption{Multi-bands light curves of 2MASS~J12184187-0609123 obtained in 2024, May 07th. The two black downward arrows mark the start and end of our time-resolved spectroscopic monitors. The vertical dashed 
line shows the trigger time of the corresponding X-ray transient triggered by the EP/WXT. 
\it Inset panel: \rm Modeling of the late $r-$band light curve by an exponential decay. See the main text for the details.   
\label{fig:general}}
\end{figure*}

\subsection{Differential H$\alpha$ Line Profile}

In order to investigate the H$\alpha$ line emission, differential spectra are obtained by 
subtracting the quiescent spectrum taken about one month after the trigger. Taking into
account the fact that the continuum during the spectroscopic observations almost returns back the quiescent level as 
shown in Figure 4,
all the spectra taken in 2024, May 07 are scaled to have a common continuum flux within the 
wavelength range 6650\AA-6800\AA\ by using the quiescent spectrum as a reference. The 
differential spectral sequence, along with the quiescent spectrum, are displayed 
in the left panel in Figure 5.

\begin{figure*}[ht!]
\plotone{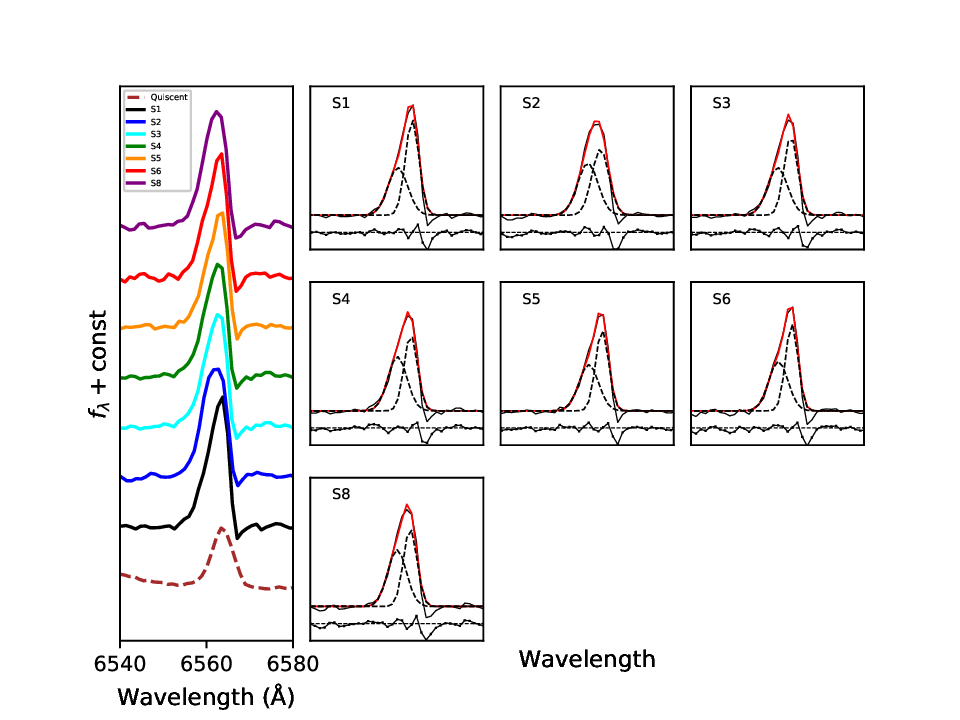}
\caption{\it Panel left: \rm Differential profile sequence of the H$\alpha$ emission-line (the 
solid lines)
sorted with time from top to bottom, along 
with the line profile obtained from the quiescent spectrum (the dashed line). See main text for the details of the generation of the differential line profile.  The spectra are shifted vertically by an arbitrary amount for clarity. 
\it Panels right: \rm  Modeling of the H$\alpha$ differential profile by a linear combination of two Gaussian functions. 
The modeled local continuum has already been removed from the original differential spectrum.  In each panel, the
observed and modeled line profiles are plotted by the black and red solid lines,
respectively. Each Gaussian function is shown by a dashed line. The curves underneath each line spectrum present
the residuals between the observed and modeled profiles.
\label{fig:general}}
\end{figure*}

Two facts can be learned from the figure. At first, there is an excess of H$\alpha$ 
line emission in all the differential spectra, which strongly suggests a flare activity 
occurring on the object and reinforces conclusion that the stellar activity is the 
origin of both X-ray transients detected by EP/WXT. Secondly, although the H$\alpha$ line emission level seems to be stable, there is a significant H$\alpha$ blue wing in all the 7 differential spectra. In contrast, 
the line profile is symmetric for the quiescent spectrum.

To quantify the H$\alpha$ line emission, 
we model its line profile on the differential spectra by 
a linear combination of two Gaussian functions by the IRAF/SPECFIT task (Kriss 1994), 
which is illustrated by the right small panels, denoted by S1-S8, in Figure 5. The results of the line profile 
modeling is tabulated in Table 3. Columns (2) and (3) list the measured line flux of the 
central and blueshifted components, respectively. The listed uncertainties  
corresponds to 1$\sigma$ significance level due to the modeling.
Columns (4) is the measured line width of the blueshifted H$\alpha$ component, after a correction 
of the instrumental profile $\mathrm{FWHM=\sqrt{FWHM_{obs}^2-FWHM_{inst}^2}}$.
The line shifts of the blueshifted component $\Delta\upsilon_{\mathrm{b}}$
are tabulated in Column (5). $\Delta\upsilon_{\mathrm{b}}=c(\lambda_{\mathrm{b}}-\lambda_{\mathrm{n}})/\lambda_0$, where $\lambda_{\mathrm{b}}$ ($\lambda_{\mathrm{n}}$) is the 
the measured center wavelength of the blueshifted central (central) component, and $\lambda_0$ 
the rest-frame wavelength in vacuum. The typical uncertainty of $\mathrm{FWHM}$ and 
$\Delta\upsilon_{\mathrm{b}}$ is $\sim10\mathrm{km\ s^{-1}}$, which is much smaller than the 
measured values. Column (6) shows the maximum projected velocity
$V_\mathrm{max}$ measured from the observed spectrum at the
position where the H$\alpha$ high-velocity blue wing merges with the
continuum, after applying the instrumental profile correction.

\begin{table*}[h!]
\centering
\caption{Results of Spectral Measurements and Analysis.}
\label{tab:decimal}
\begin{tabular}{cccccc}
\tablewidth{0pt}
\hline
\hline
ID & $f\mathrm{(H\alpha_n)}$ & $f\mathrm{(H\alpha_b)}$ &  $\mathrm{FWHM(H\alpha)}$ & $\Delta\upsilon(\mathrm{H\alpha_b})$ & $V_{\mathrm{max}}$\\
 &  \multicolumn{2}{c}{$(\mathrm{10^{-15}erg\ s^{-1}\ cm^{-2}})$} & \multicolumn{3}{c}{($\mathrm{km\ s^{-1}})$} \\
(1)  &   (2) & (3) & (4) & (5) & (6) \\
\hline
1 &  $86.6\pm1.2$  & $63.5\pm2.6$  & 200 &  -150  & -240 \\
2 &  $75.6\pm8.0$  & $71.6\pm7.0$  & 210 &  -130  & -250 \\
3 &  $74.7\pm3.7$  & $65.3\pm4.1$  & 210 &  -140  & -240 \\
4 &  $67.9\pm2.0$  & $69.8\pm1.4$  & 190 &  -140  & -230 \\
5 &  $70.4\pm2.0$  & $61.1\pm2.8$  & 200 &  -140  & -230 \\
6 &  $72.5\pm1.6$  & $64.2\pm3.2$  & 190 &  -140  & -180 \\
8 &  $70.8\pm2.5$  & $72.0\pm3.3$  & 190 &  -140  & -210 \\
\hline
\hline
\end{tabular}
\end{table*}


\section{Discussion}

The photometrical and spectroscopic follow-up observations in optical bands enable us to identify the X-ray transient triggered by EP/WXT in 
2024 May 07 is resulted from a superflare of the M-dwarf 2MASS~J12184187-060912. 
The flaring energy released in X-ray 
is estimated to be $\sim10^{34}\ \mathrm{erg}$.
The H$\alpha$ emission line profile 
obtained at about 7 hours after the trigger shows a strong blue asymmetry
with a maximum velocity of the blue wing of 
$\approx200-250\ \mathrm{km\ s^{-1}}$.

A CME origin of the observed blueshifted H$\alpha$ emission is 
argued against due to its small measured $V_{\mathrm{max}}\sim200\ \mathrm{km\ s^{-1}}$.
In fact, 
given the formula of $\upsilon_{\mathrm{esp}}=630(M_\star/M_\odot)^{1/2}(R_\star/R_\odot)^{-1/2}\ \mathrm{km\ s^{-1}}$, the value of $\upsilon_{\mathrm{esp}}$ is estimated to be 
590$\mathrm{km\ s^{-1}}$ for the object 2MASS~J12184187-0609123.

\subsection{Chromospheric Evaporation}
The blueshifted H$\alpha$ emission can be possibly explained by the chromospheric temperature (cool)
upflow associated with
chromospheric evaporation (e.g., Canfield et al. 1990; Gunn
et al. 1994; Berdyugina et al. 1999; Tei et al. 2018; Li 2019). Although it is not commonly observed, 
similar blueshifted H$\alpha$ emission possibly caused by chromospheric evaporation has been 
infrequently reported in previous studies (e.g., Koller et al. 2021; Wang et al. 2022; Cao \& Gu 2024).

In the evaporation scenario, the chromospheric plasma is heated to a very high
temperature rapidly through Coulomb collisions of the electrons accelerated by the 
energy released in the magnetic reconnection (e.g., Fisher
et al. 1985; Innes et al. 1997; Li 2019; Yan et al. 2021; Fletcher et al. 2011; Tan et al. 2020; 
Chen et al. 2020). The heating results in an overpressure in the chromosphere, 
which then pushes the plasma upward (e.g., Fisher et al. 1985; Teriaca et al. 2003; Zhang et al. 2006b; Brosius \& Daw 2015; Tian \& Chen 2018)
or downward (e.g., Kamio
et al. 2005; Zhang et al. 2006a; Libbrecht et al. 2019; Graham et al. 2020).
By a careful examination of the 
differential H$\alpha$ line profile shown in Figure 5, one can see a weak and sharp absorption with a velocity of $\sim160\mathrm{km\ s^{-1}}$ 
at the red wing in all the seven spectra. 
By analyzing the time-resolved spectroscopy of active M-dwarf EV Lac,  
Honda et al. (2018, and references therein) suggests that the observed redshifted
absorption in EV Lac might be caused by downward plasma.
In the case of a ``explosive evaporation'', 
a fast upward motion with a velocity of hundreds of kilometers per second can be resulted 
by an electron beaming flux $\geq3\times10^{10}\ \mathrm{erg\ cm^{-2}\ s^{-1}}$ (e.g.,
Milligan et al. 2006b; Brosius \& Inglis 2017; Li et al. 2017).

\subsection{Mass of the Moving Plasma}

In the chromospheric evaporation scenario, we then estimate the mass of the moving plasma
from the blueshifted H$\alpha$ emission through the traditional method $M_{\mathrm{gas}}\geq N_{\mathrm{tot}}Vm_{\mathrm{H}}$ (e.g., Houdebine et al. 1990), where $N_{\mathrm{tot}}$ is the
number density of hydrogen atoms, $m_{\mathrm{H}}$ the mass of the hydrogen atom, and $V$ the total 
volume that can be determined from the line luminosity $L_{ji}$.
After involving $L_{ji}=N_jA_{ji}h\nu_{ji}VP_{\mathrm{esc}}$, 
the mass of the moving plasma $M_{\mathrm{PL}}$ is therefore estimated as
\begin{equation}
 M_{\mathrm{PL}}\geq\frac{4\pi d^2f_{\mathrm{line}}m_{\mathrm{H}}}{A_{ji}h\nu_{ji}VP_{\mathrm{esc}}}\frac{N_{\mathrm{tot}}}{N_j}
\end{equation}
where $N_j$ is the number
density of hydrogen atoms at excited level $j$, $A_{ji}$ the Einstein
coefficient for a spontaneous decay from level $j$ to $i$, $P_{\mathrm{esc}}$
the escape probability, $d$ the distance and $f_{\mathrm{line}}$ the observed line flux.
With the above equation and a typical value of $P_{\mathrm{esc}}=0.5$, the measured line flux of the blueshifted H$\alpha$ emission yields 
$M_{\mathrm{PL}} = 1.2\times10^{18}$g, after transforming the H$\alpha$ line flux to that of 
H$\gamma$ by assuming a Balmer decrement of three (Butler et al. 1988).
For H$\gamma$ line, we have $A_{52}=2.53\times10^6\ \mathrm{s^{-1}}$ (Wiese \& Fuhr 2009) 
and $N_{\mathrm{tot}}/N_5= 2\times 10^9$ estimated from nonlocal thermal equilibrium modeling by 
Houdebine \& Doyle (1994a, 1994b).

\subsection{Prominence Eruption}

The blushifted H$\alpha$ emission with a velocity of $\approx150\ \mathrm{km\ s^{-1}}$
observed in the object could be aternatively 
explained by the prominence eruption that has been detected in the decay phase of 
the flares in other stars and the Sun (e.g., Kurokawa et al. 1987). 
In addition, due to a lack of significant white-light
emission caused by chromospheric condensation, Inoue et al. (2024a) recently 
proposed that the 
prominence eruption is a possible explaination of the blueshifted H$\alpha$ emission at 
velocity of $\sim100\ \mathrm{km\ s^{-1}}$ detected at one hour after the flare peak 
in EV Lac. 
Based on a Sun-as-star analysis, Otsu et al. (2022) shows that 
the off-limb prominence eruption is able to produce 
both blueshifted and redshifted H$\alpha$ emission in
solar-like stars.

We estimate the mass of the prominence by following the method adopted in 
Maehara et al. (2021) and Inoue et al. (2023):  
$M_{\mathrm{p}}\approx m_{\mathrm{H}}n_{\mathrm{H}}A_{\mathrm{p}}D$, where $m_{\mathrm{H}}$
is the mass of the hydrogen atom, $n_{\mathrm{H}}$ the hydrogen atom, 
$D$ the geometrical 
thickness, and $A_{\mathrm{p}}$ the area of the region emitting H$\alpha$. 
The area can be estimated from the blueshifted H$\alpha$ line luminosity through the 
integration
\begin{equation}
  L_{\mathrm{H\alpha}} = \iint F_{\mathrm{H\alpha}}\mathrm{d}A\mathrm{d}\Omega=2\pi A_{\mathrm{p}}F_{\mathrm{H\alpha}}
\end{equation}
where $F_{\mathrm{H\alpha}}$ is the prominence H$\alpha$ emission in unit of 
$\mathrm{erg\ s^{-1}\ cm^{-2}\ sr^{-1}}$, and depends on the optical depth. 
We finally have 
\begin{equation}
  M_{\mathrm{p}}\approx 2m_{\mathrm{H}}\bigg(\frac{n_{\mathrm{H}}}{n_{\mathrm{e}}}\bigg)n_{\mathrm{e}}^{-1}\frac{d^2f_{\mathrm{H\alpha}}}{F_{\mathrm{H\alpha}}}\times EM
\end{equation}
where $n_{\mathrm{H}}/n_{\mathrm{e}}=2.13-5.88$ for a prominence (Notsu et al. 2024),
$EM=n_{\mathrm{e}}^2D$ is the emission measurement, $d$ the distance and 
$f_{\mathrm{H\alpha}}$ the measured H$\alpha$ line flux. The typical 
electron density of a solar prominence is found to range from 
$10^{10}\ \mathrm{cm^{-3}}$ to $10^{11.5}\ \mathrm{cm^{-3}}$ (Hirayama 1986).

The optical depth of the prominence H$\alpha$ emission is assumed to be 
$0.1<\tau<100$ by following Inoue et al. (2023). The 
NLTE solar prominence model (Heinzel et al. 1994) gives that 
$F_{\mathrm{H\alpha}}\sim10^{4}\ \mathrm{erg\ s^{-1}\ cm^{-2}\ sr^{-1}}$ and 
$EM\sim10^{28}\ \mathrm{cm^{-5}}$ in the case of $\tau=0.1$. 
In the case of $\tau=100$, the corresponding values 
are $F_{\mathrm{H\alpha}}\sim10^{6}\ \mathrm{erg\ s^{-1}\ cm^{-2}\ sr^{-1}}$
and $EM\sim10^{31}\ \mathrm{cm^{-5}}$. With these values, the 
measured blushifted H$\alpha$ line flux list in the Column (3) in Table 3 
finally leads to an estimation of prominence mass of 
$7\times10^{15}\ \mathrm{g}<M_{\mathrm{p}}<7\times10^{18}\ \mathrm{g}$ 
for the objects. 

This estimated mass is comparable with the CME mass of M-dwarfs complied in 
Moschou et al. (2019) and recently measured by Wang et al. (2022) and Wang (2023).

\subsection{Energy Budget and Possible Accompanying CME}

By adopting a bolometric correction of 
$L_{\mathrm{X}}/L_{\mathrm{bol}}=0.01$ determined for not only the Sun (e.g., Kretzschmar 2011; Emslie et al. 2012), but also stars (e.g., Wang et al. 2022), the total flaring energy is estimated to be
$E_{\mathrm{bol}}=3\times10^{36}$ and $5\times10^{36}\ \mathrm{erg}$ for 
the Flare 1 and 2, respectively. Alternatively, a lower limit of the bolometric
flaring energy of $E_{\mathrm{bol}}=1.4\times10^{32}$erg  
can be inferred from our modeling of the optical light curve at lat epoch, 
by adopting a bolometric correction of $E_{\mathrm{bol}}/E_g=12$, when a black body
with a temperature of $10^4$K (e.g., Gizis et al. 2013; Kowalski et al.
2013; Paudel et al. 2019; Fleming et al. 2022; Murray et al. 2022) is assumed.

The observations of the Sun reveal a linkage between solar flares
and associated 
coronal mass ejection (CME). That is: the more powerful a solar flare, more
intense the associated CME will be (e.g., Yashiro et al. 2008; Aarnio et al. 2011 , 2012; Webb \& Howard 2012). This linkage seems to be valid 
for the stellar flares (e.g., Moschou et al. 2019; Wang et al. 2022, 2023),
although undoubted detection of stellar CMEs is still an open issue at current stage
(Leitzinger \& Odert 2022). 

Although both the Flare 1 and 2 are powerful enough, our spectroscopic observations were
too late to allow us to investigate the possible associated CME, because 
the Flare 1 occurring in May 07 was triggered in our afternoon.   
We instead estimate the possible CME velocity $V_{\mathrm{CME}}$ from the 
X-ray emission detected by EP/WXT through the empirical relationship 
$\log(V_{\mathrm{CME}}/\mathrm{km\ s^{-1}})=(0.20\pm0.08)\log(F_{\mathrm{p}}/\mathrm{W\ m^{-2}})+(3.83\pm0.38)$ established for 
solar CMEs (Salas-Matamoros \& Klein 2015; Moon et al. 2002; 
Yashiro \& Gopalswamy 2009), where $F_{\mathrm{p}}$ is the 
peak soft X-ray flux received at a distance of 1AU. The calculated 
X-ray peak luminosity therefore predicts $V_{\mathrm{CME}}\approx7200$ and 
$\approx7800\ \mathrm{km\ s^{-1}}$ for the Flare 1 and 2, respectively.

\section{Conclusions}
Based on our photometrical and spectroscopic follow-up observations in optical bands, we identify 
an X-ray transient triggered by EP/WXT as a stellar 
flare whose bolometric energy is estimated to be $\sim10^{36}\ \mathrm{erg}$ from the X-ray light curve. Evident blue 
wing with a maximum velocity of $200-250\ \mathrm{km\ s^{-1}}$ is detected in 
the H$\alpha$ emission-line
profile taken at about 7 hrs after the trigger.
By explaining the blue wing by the chromospheric temperature (cool) upflow associated
with chromospheric evaporation,
the mass of the evaporating plasma is estimated to be $1.2\times10^{18}$g. 
This blue wing could be alternatively explained by a prominence eruption with an estimated mass of 
$7\times10^{15}\mathrm{g}<M_{\mathrm{p}}<7\times10^{18}\mathrm{g}$.


\acknowledgments

We thank the anonymous referee for the helpful comments that improved our study significantly.
This study is supported by the Strategic
Pioneer Program on Space Science, Chinese Academy of
Sciences, grants XDB0550401. JW is supported by
the National Natural Science Foundation of China (under grants 12173009) and the Natural Science Foundation of Guangxi (2020GXNSFDA238018). H.Y. acknowledges the support by NSFC (grant No. 12103061).
EWL is supported by 
the National Postdoctoral Program for Innovative Talents (grant No. 12133003) and 
the Guangxi Talent Program(``Highland of Innovation Talents''). 
We acknowledge the support of the staff of the Xinglong 2.16m telescope, SVOM/C-GFT and SVOM/GWAC. 
This work was partially supported by the Open Project Program of the Key Laboratory of Optical Astronomy, National Astronomical Observatories, Chinese Academy of Sciences. SVOM/C-GFT is supported by the 
SVOM project, a mission in the Strategic Priority Program on Space Science of CAS.
This study used the NASA/IPAC Extragalactic Database (NED), which is operated by the Jet Propulsion Laboratory, California Institute of Technology.

\vspace{5mm}
\facilities{Einstein Probe (Yuan et al. 2022), Beijing:2.16m (Fan et al. 2016), SVOM/CGFT, GWAC/F60A}
\software{IRAF (Tody 1986, 1992), MATPLOTLIB (Hunter 2007), Heasoft,  Xspec (Arnaud 1996)}
%




\end{document}